\documentclass[conference]{IEEEtran}
\IEEEoverridecommandlockouts
\usepackage{cite}
\usepackage{amsmath,amssymb,amsfonts}
\usepackage{algorithmic}
\usepackage{graphicx}
\usepackage{textcomp}
\usepackage{cuted}
\usepackage{xcolor}
\def\BibTeX{{\rm B\kern-.05em{\sc i\kern-.025em b}\kern-.08em
    T\kern-.1667em\lower.7ex\hbox{E}\kern-.125emX}}
    
\usepackage{acronym}
\usepackage{comment}
\usepackage[capitalise,noabbrev]{cleveref}
\Crefformat{figure}{#2Fig.~#1#3}
\Crefformat{equation}{#2(#1)#3}

\usepackage[detect-all,range-phrase=--,per-mode=symbol,range-units=single,list-units=single]{siunitx}
\DeclareSIUnit\dBm{dBm}
\DeclareSIUnit\dB{dB}
\DeclareSIUnit\dBi{dBi}
\acrodef{FoV}{field of view}    
\acrodef{SDR}{software defined radio}
\acrodef{MIMO}{multiple input multiple output}
\acrodef{UT}{user terminal}
\acrodef{CSI}{channel state information}
\acrodef{DRA}{direct radiating active antenna array}

\begin{document}
\bstctlcite{IEEEexample:BSTcontrol}

\title{Demonstrator Testbed for Effective Precoding in MEO Multibeam Satellites       
\thanks{This work has been supported by the European Space Agency under the project number 4000122451/18/NL/NR "Live Satellite Precoding Demonstration - CCN: MEO case (LiveSatPreDem)." 
(Opinions, interpretations, recommendations, and conclusions presented in this paper are those of the authors and are not necessarily endorsed by the European Space
Agency).}
}

\author{\IEEEauthorblockN{
Jorge L. Gonz\'alez-Rios\textsuperscript{*}, 
Liz Mart\'inez Marrero,
Juan Duncan\textsuperscript{*}, 
Luis M. Garc\'es-Socarr\'as,\\\
Raudel Cuiman Marquez,
Juan A. V\'asquez Peralvo, 
Jevgenij Krivochiza,
Symeon Chatzinotas,
Björn Ottersten 
}
\IEEEauthorblockA{
SnT, University of Luxembourg, Luxembourg \\
\textsuperscript{*}Corresponding authors: \{jorge.gonzalez,juan.duncan\}@uni.lu}
}

\maketitle

\begin{abstract}
The use of communication satellites in medium Earth orbit (MEO) is foreseen to provide quasi-global broadband Internet connectivity in the coming networking ecosystems. Multi-user multiple-input single-output (MU-MISO) digital signal processing techniques, such as precoding, emerge as appealing technological enablers in the forward link of multi-beam satellite systems operating in full frequency reuse (FFR). However, the orbit dynamics of MEO satellites pose additional challenges that must be carefully evaluated and addressed. This work presents the design of an in-lab testbed based on software-defined radio (SDR) platforms and the corresponding adaptations required for efficient precoding in a MEO scenario. The setup incorporates a precise orbit model and the radiation pattern of a custom-designed direct radiating array (DRA). We analyze the main impairments affecting precoding performance, including Doppler shifts and payload phase noise, and propose a synchronization loop to mitigate these effects. Preliminary experimental results validate the feasibility and effectiveness of the proposed solution.

\end{abstract}


\section{Introduction}

The use of communication satellites in medium Earth orbit (MEO) for quasi-global broadband Internet connectivity is likely to be instrumental within the sixth generation (6G) networking ecosystem~\cite{9268898}. The current rising need for broadband and trunking traffic has prompted the pursuit of technologically innovative approaches that can supply additional spectrum to next-generation systems. MEO satellites offer lower latency than geostationary orbit (GEO) ones, with better network scalability while maintaining link reliability~\cite{7896575}. 

Multi-user multiple-input single-output (MU-MISO) digital signal processing techniques, like precoding~\cite{9351985, ha2024usercentric,eappen2025livesat16},  emerge as appealing technological enablers in the forward (FWD) link of multi-beam satellite systems operating in full frequency reuse (FFR). 
Precoding will empower future MEO missions by providing additional flexibility and capacity, and can simplify the design of future satellite payloads, moving some of the complexity toward the ground segment, which can afford a higher processing burden on the gateway (GW) side.

To evaluate the feasibility of precoding techniques in communication satellites, our group initiated the LiveSatPredem (Live Satellite Precoding Demonstration) project in 2018.
Through live demonstrations and laboratory-controlled end-to-end emulations, this activity focuses on precoding for the forward link of multi-beam satellite systems operating in FFR~\cite{live_jevgenij2021,9771918,eappen2025livesat16}.
Nevertheless, the precoding technique has been more used in GEO than in MEO satellite systems \cite{Storek2020,live_jevgenij2021}. 
This situation has changed with the increasing interest in MEO satellites, with different research focusing on the application of precoding in this kind of orbit \cite{Vazquez2021,Vidal2020a,vidal_system_2021}. 
However, the published works have been solely relying on simulations for validating the proposed solutions and approaches.

In this context, this work presents the design of an in-lab testbed based on software-defined radio (SDR) and the corresponding adaptations for precoding in a MEO scenario.
The main contributions of this article are as follows:
(i) Developing a realistic MEO multibeam satellite scenario by integrating precise orbital dynamics and the radiation pattern of a custom-designed direct radiating array (DRA). 
(ii) Designing and implementing a phase compensation loop capable of mitigating fast differential phase variations, particularly those induced by uplink Doppler shifts and payload phase noise. (iii) Upgrading an existing SDR-based MIMO satellite channel emulator to reproduce MEO-specific impairments. 
(iv) Experimentally assessing the impact of individual and combined impairments on precoding performance and demonstrating the effectiveness of the proposed mitigation approach.


\section{System Model}
\label{sec:system_model}
We consider a MEO satellite equipped with a DRA capable of generating $N$ beams serving $K\leq N$ single-antenna UTs.
We collect in $\mathrm{\mathbf{h}}_k \in \mathbb{C}^{N \times 1}$ the complex (i.e., magnitude and phase) coefficients of the frequency-flat slow fading channels between the beams generated at the GW and the $k\text{-th}$ UT. 
At a given symbol period, independent data symbols $\{s_k\}_{k=1}^K$ are to be transmitted to the UTs, where $s_k$ denotes the symbol intended for the $k\text{-th}$ user. 
Under the above assumptions, the received vector containing the symbol-sampled complex baseband signals of all $K$ UTs can be modeled as
\vspace{-3pt}
\begin{equation}\label{eq:sys}
\mathrm{\mathbf{r}} = \mathrm{\mathbf{HWs}}+\mathrm{\mathbf{z}},
\end{equation}
where $\mathrm{\mathbf{H}}=[\mathrm{\mathbf{h}}_1\,\cdots\,\mathrm{\mathbf{h}}_K]^\mathrm{T}$ denotes the $K \times N$ complex-valued channel matrix, $\mathrm{\mathbf{W}}$ stands for the $N \times K$ precoding matrix, $\mathrm{\mathbf{s}} = [s_1\, \cdots\,s_K]^\mathrm{T}$ is a $K \times 1$ complex-valued vector containing the UTs' intended modulated symbols, and $\mathrm{\mathbf{z}}$ collects the independent complex zero-mean additive white Gaussian noise (AWGN) components at the UTs' receivers. 

The channel matrix collecting the complex channel coefficients for all the $K$ UTs can be written as
\begin{equation}
\mathrm{\mathbf{H}}(t) = \begin{bmatrix} |h_{11}(t)|{\rm e}^{\mathrm{j}\psi_{11}(t)} & \cdots & |h_{1N}(t)|{\rm e}^{\mathrm{j}\psi_{1N}(t)} \\
|h_{21}(t)|{\rm e}^{\mathrm{j}\psi_{21}(t)} & \cdots & |h_{2N}(t)|{\rm e}^{\mathrm{j}\psi_{2N}(t)} \\
\vdots & \vdots  & \vdots \\
|h_{K1}(t)|{\rm e}^{\mathrm{j}\psi_{K1}(t)} & \cdots & |h_{KN}(t)|{\rm e}^{\mathrm{j}\psi_{KN}(t)}
\end{bmatrix},
\label{eq:H}
\end{equation}
where $h_{kj}$ denotes the channel coefficient between the $k\text{-th}$ UT and the $j\text{-th}$ generated beam in the transmit antenna, for any $k\in\{1,2,...,K\}$ and $j\in\{1,2,...,N\}$, and $|h_{k,j}|$ and $\psi_{k,j}$ respectively denote its magnitude and phase. The time-varing terms $|h_{kj}(t)|$ represent the continuously changing elevation angle between the satellite and the user terminals due to the movement of the satellite. 


In addition, we have to include a time-varying matrix $\mathbf{\Gamma} (t) \triangleq \text{diag}(e^{\mathrm{j}\epsilon_{1}(t)},e^{\mathrm{j}\epsilon_{2}(t)},...,e^{\mathrm{j}\epsilon_{K}(t)})$ with the phase variations due to Doppler effect  
The complete channel matrix can then be written as
\begin{equation}\label{eq:Hmeo}
   \mathrm{\mathbf{H}}_{\text{MEO}}(t) = \mathbf{\Gamma} (t) \mathrm{\mathbf{H}}(t)
\end{equation}

The time-varying gain and the phase variations in the channel matrix are addressed by the precoding design considered in this paper. These two impairments depend on the orbit described by the satellite and the ground location of the UTs.

\subsection{Satellite Orbit Model}
\label{sec:orbit}

The performance of the forward channel depends on time-varying propagation effects, including those due to the satellite's orbital motion. 
We obtained these parameters using MATLAB for a MEO satellite system, although the same method can be used for other non-geostationary orbits (NGSOs), such as LEO.
The input to our model is the TLE data for the O3B FM5 satellite read from the Celestrak website\footnote{https://celestrak.org/} and the on-board antenna described in the next subsection.
We considered a GW located in Dakar, and four UTs located nearby.
The coordinates for the ground stations and the carrier frequencies are included in \Cref{tab:scenario_ground,tab:scenario_freq}, respectively.

\begin{table}[h!]
\centering
\caption{Ground station coordinates}
\begin{tabular}{|c|c|c|}
\hline
\textbf{Ground station} & \textbf{Latitude} & \textbf{Longitude} \\ \hline
GW             & 14.743   & -17.491   \\ \hline
UT0            & 12.935   & -19.349   \\ \hline
UT1            & 12.935   & -17.491   \\ \hline
UT2            & 14.743   & -19.349   \\ \hline
UT3            & 14.743   & -17.491   \\ \hline
\end{tabular}
\label{tab:scenario_ground}
\end{table}

\begin{table}[h!]
\centering
\caption{Frequency Plan}
\begin{tabular}{|c|c|c|c|c|c|}
\hline
\textbf{Beam} & Uplink0 & Uplink1 & Uplink2 & Uplink3 & Downlink \\ \hline
\textbf{\begin{tabular}[c]{@{}c@{}}Carrier Freq.\\ (GHz)\end{tabular}} & 47.2 & 47.7 & 48.2 & 48.7 & 20.0 \\ \hline
\end{tabular}
\label{tab:scenario_freq}
\end{table}
\vspace{-6pt}

\subsection{Antenna Design}
\label{sec:Antenna_design}
A DRA was selected for a MEO satellite link in the 17.7--20.2~GHz band, aiming to generate a $\approx$150,000~km$^2$ spot beam with right- and left-hand circular polarization, $2^\circ$ beamwidth, and a $50\times50$ element configuration. 
The antenna elements are fed through a microstrip network with a $90^\circ$ phase shift, coupled via a square aperture to excite orthogonal modes for circular polarization, and optimized with foam layers to improve impedance bandwidth and axial ratio.
Simulations in CST confirmed compliance with the reflection coefficient, axial ratio, and gain requirements.
More details on the antenna element design can be found in \cite{vazquez2022circular_antenna}.

\begin{figure}[h!]
\centerline{\includegraphics[width=0.75\linewidth]{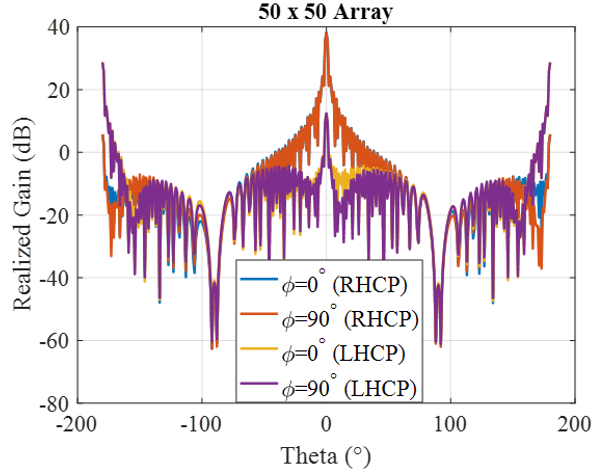}}
\caption{Co-polar and cross-polar full array radiation pattern.}
\label{Figure_ant_5}
\end{figure}

The array, with $\lambda_0/2$ element spacing and overall size of $25\lambda_0\times25\lambda_0$, was analyzed using the array factor formulation, combining the element pattern and array geometry to obtain the total gain and directivity. The resulting radiation pattern (Fig.~\ref{Figure_ant_5}) shows a cross-polarization component about 20~dB below the co-polar maximum, a $2^\circ$ beamwidth at boresight, and similar principal cuts, ensuring the pencil beam shape required for beam steering.

\subsection{Scenario Definition}
\label{sec:scenario}

\begin{figure}[h!]
    \centering
    \includegraphics[width=0.8\linewidth]{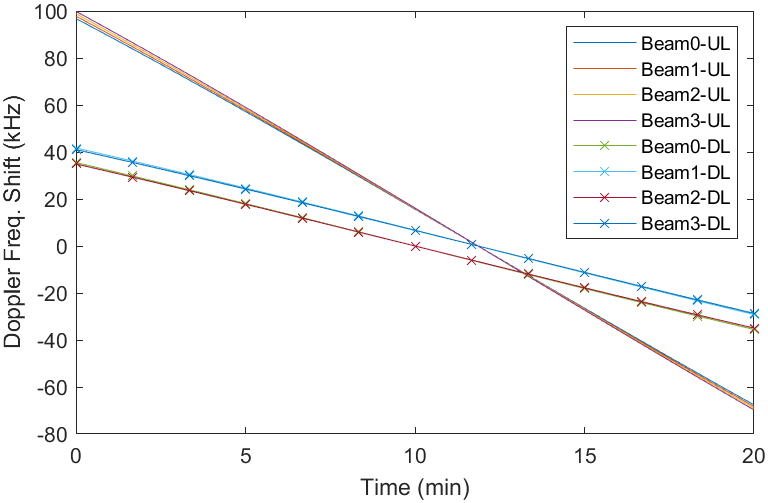}
    \caption{Doppler shift for each beam in uplink (UL) and downlink (DL).}
    \label{fig:doppler}
\end{figure}

\begin{figure}[h!]
    \centering
    \includegraphics[width=0.8\linewidth]{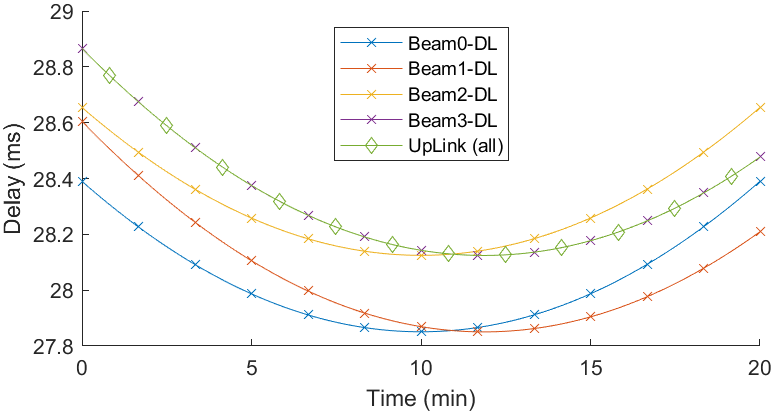}
    \caption{Delay for each beam in uplink and downlink (DL).}
    \label{fig:delay}
\end{figure}

\begin{figure}[h!]
    \centering
    \includegraphics[width=0.8\linewidth]{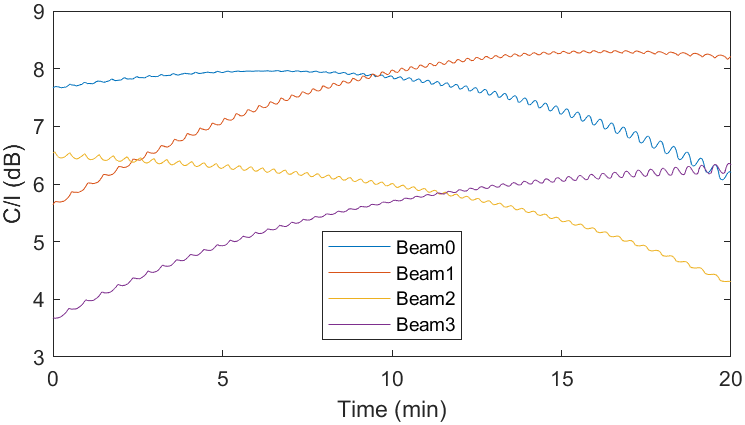}
    \caption{Nominal carrier-to-interference ratio (C/I)}
    \label{fig:c_over_i}
\end{figure}

\Cref{fig:doppler,fig:delay,fig:c_over_i} depicts the calculated orbital and channel parameters for the selected scenario, considering 20 minutes around the zenith.
In the case of the Doppler shift (c.f. \Cref{fig:doppler}), in the experiments we apply a scaled version of the original values, considering that the GW can estimate the Doppler shift with a precision of +/-1~kKz \cite{ccsds2021rf_rec} and apply a precompensation (we do not assume perfect compensation though).
The carrier-to-interference-ratio (C/I) in \Cref{fig:c_over_i} is calculated from the 16 (4$\times$4) complex channel coefficients and provided here as a compact representation of the channel, although the testbed uses the full time-varying matrix as an input.
The ripple in the C/I curves arises from the satellite's movement, which is not perfectly elliptical and is affected by the actual orbital perturbations, combined with the effects of the changes on the pointing angle on the DRA radiation pattern.

\begin{figure}[!htbp]
    \centering
    \includegraphics[width=0.7\linewidth]{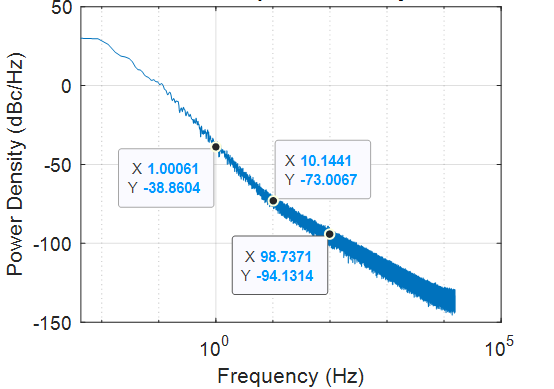}
    \caption{Phase noise power spectral density for each beam.}
    \label{fig:phase_noise}
\end{figure}

Moreover, \Cref{fig:phase_noise} shows the power spectral density of the phase noise applied to each beam (fully uncorrelated), following the model described in \cite{marrero2024sync}.

\section{Precoding with Phase Compensation Design}
\label{sec:precoding}

\begin{figure*}[htbp]
\centerline{\includegraphics[width=0.65\textwidth]{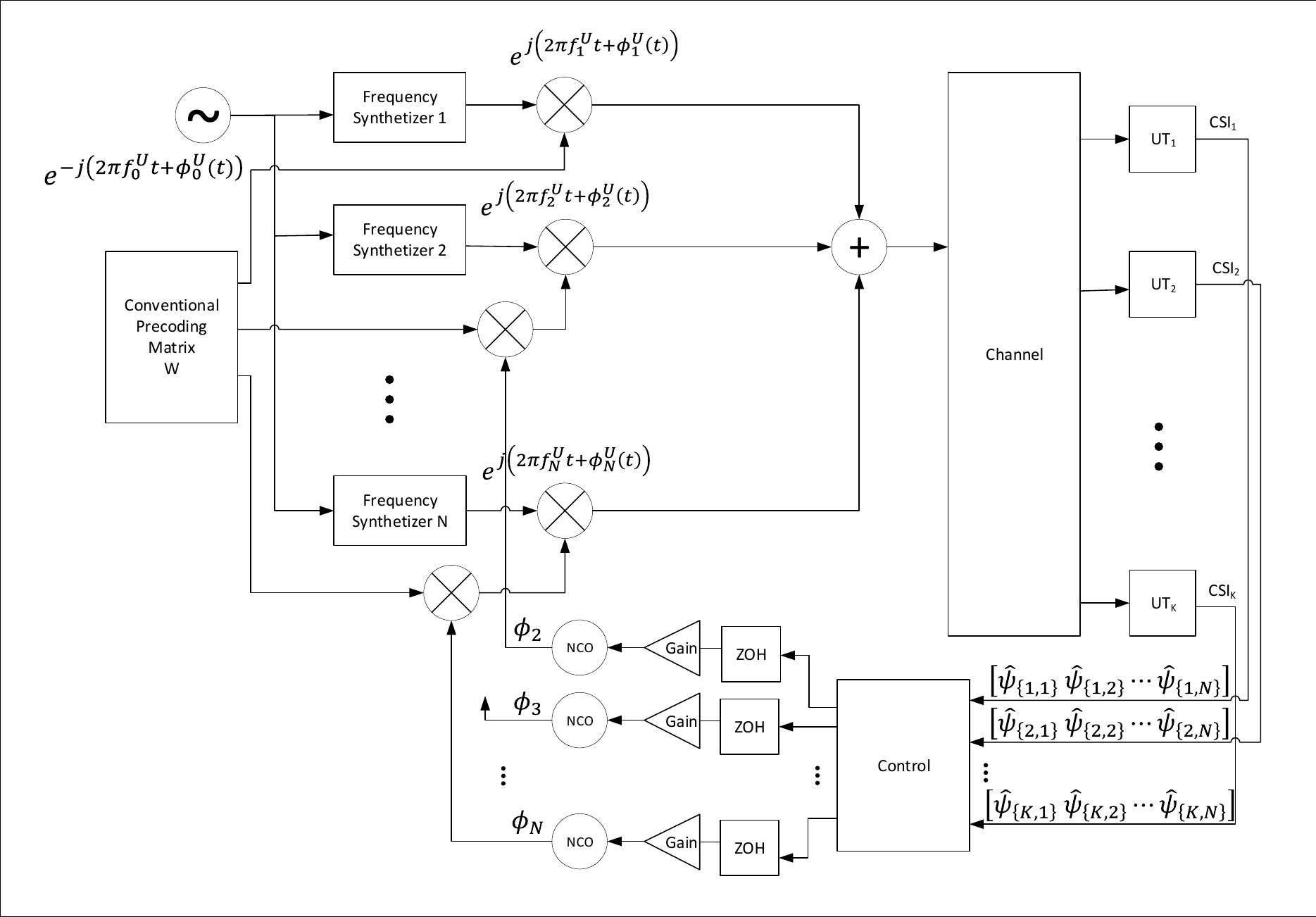}}
\caption{Differential phase-error compensation loop.}
\label{fig::synch_loop}
\end{figure*}

Conventional precoding assumes perfect beam phase control, but practical systems suffer from fast phase variations introduced by the phase noise of local oscillators (LOs) and Doppler shift. 
It was formally demonstrated in \cite{Marrero2022} that precoding techniques performance is only affected by the phase errors in the uplink channel. 
That is, independent phase rotations per beam, such as the differential Doppler shift in the uplink due to frequency-division multiplexing (FDM), the phase noise of the transponder's LO. 
These effects cause rapid differential phase errors that the conventional, slow-updating precoding loop cannot fully compensate.
To address this, we implement a second-order, fast-response compensation loop operating in a sample-based mode, enabling more effective mitigation than the symbol/frame-based updates used in standard precoding.
The solution is represented in Fig.~\ref{fig::synch_loop}, where only phase and frequency parameters are considered. In the figure, the $n$-th beam is represented by its carrier frequency $f^{\rm U}_n$, and the phase noise introduced in the uplink process is represented as $\phi^{\rm U}_n$. The Channel block includes the phase rotations due to the Doppler effect as explained in \eqref{eq:Hmeo}. The CSI estimated by the $k$-th UT has the form $[\hat\psi_{k,1} \, \hat\psi_{k,2} \, \cdots \, \hat\psi_{k,N}]$, with $\hat\psi_{k,j}=\psi_{k,j}-\psi_{k,k}$. The term $\psi_{k,j}$ includes the phase errors mentioned before. This CSI is obtained through a conventional data-aided algorithm using non-precoded pilots.  

The differential phase error is compensated at the GW by modifying the frequency offset applied to the numerically controlled oscillator (NCO) used in the digital up-conversion. Unlike conventional precoding implementations, where the differential phase compensation is updated for several symbols or frames, the phase error is compensated for each sample in our design. To this end, a compensation phase $\phi_n$ is generated by a time-invariant linear controller using the interpolated version of the received CSI phase as input. 

\section{Hardware Test-Bed Implementation}
\label{sec:hardware}

\begin{figure}[htbp!]
\centering
\includegraphics[width=.75\linewidth]{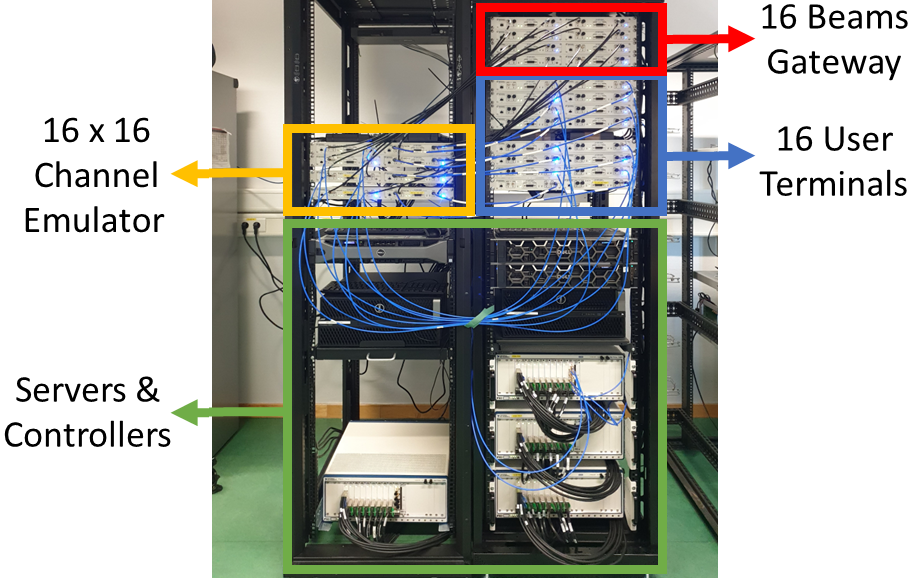}
\caption{SDR-based MIMO end-to-end satellite emulator.}
\label{fig:testbed}
\end{figure}

For the experimental validation, we employ the in-house developed MIMO end-to-end satellite emulator based on software-defined radio (SDR) platforms, shown in Fig.~\ref{fig:testbed}.
The proposed architecture consists of a multichannel GW with precoding capabilities, a MIMO satellite channel emulator (ChEm), a set of independent UTs, and a return-link emulator.

In general terms, the demonstrator can be described as follows.
The GW subsystem generates the data packets according to the DVB-S2X standard, using Superframe Format~II structure, and applies the selected precoding method.
The ChEm replicates the whole forward link chain, from the intermediate frequency (IF) input of the gateway block up-converter (BUC), toward the low-noise block down-converter (LNB) IF output at the user terminal.
It emulates the impairments present in the GW, the payload, the downlink channel, and the UTs.
The UT subsystems implement the synchronization and decoding features in the DVB-S2X-compliant receivers and perform the CSI estimation.
Finally, the return-link emulator allows each UT to send its estimated CSI to the GW.

The GW, ChEm, and UT subsystems are being implemented using a set of SDR platforms, specifically the USRP-2944R from National Instruments. 
The physical interfaces of the channel emulator with the gateway and the user terminals are provided by the interconnection of the $\text{50-}\Omega$ ports of the SDRs, employing IF modulated signals.
The SDR platforms in the GW and the channel emulator are synchronized with the same frequency and clock references. 
This eliminates any timing misalignment and additional phase noise due to USRPs' LOs, allowing precise control of the time, frequency, and phase impairments according to the implemented models.

\begin{figure*}[!htbp]
    \centering
    \includegraphics[width=.75\linewidth]{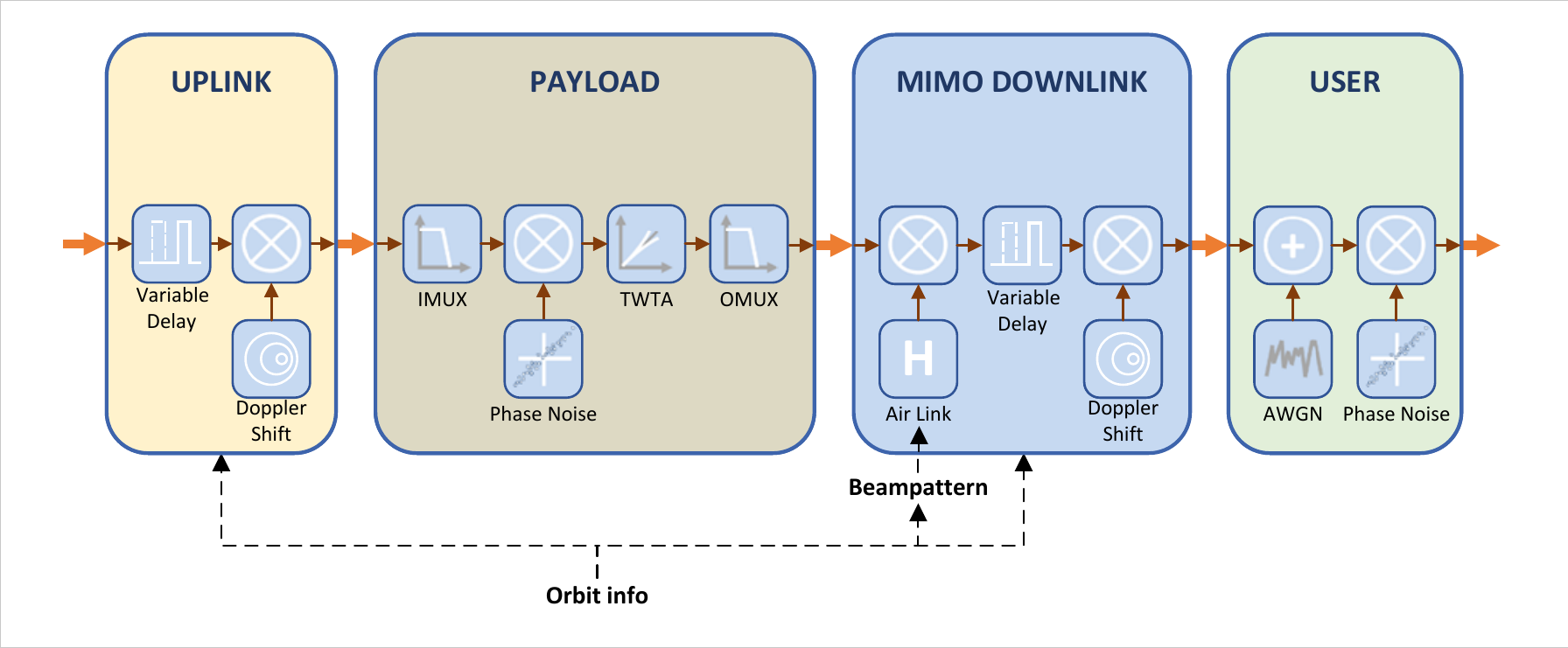}
    \caption{Block diagram of the NGSO Channel Emulator.}
    \label{fig:chem}
\end{figure*}

All the components of the system have been successfully tested considering a GEO satellite scenario~\cite{demonstrator_jc2018,live_jevgenij2021}.
This includes the use of implementations of the GW and UTs subsystems in the precoding validation over a live GEO link~\cite{live_jevgenij2021}.
The channel emulator is upgraded with the inclusion of the time-varying channel matrix in the downlink, the Doppler frequency shifts, and the delays, as shown in \Cref{fig:chem}.
These effects are obtained from the orbit and antenna models, presented in \cref{sec:system_model}.
Some of the new blocks have also been tested individually in other activities, namely the Doppler shift in \cite{gonzalez2024cgd}, and the variable delay in \cite{gonzalez2025cadsat}.
The GW subsystem incorporates the implementation of the compensation technique \cite{marrero2024sync}.

\section{Results}

This section presents the assessment of the MEO impairments' impact on the precoding scheme and the validation of the phase compensation technique. 
On top of the scenario parameters described in \Cref{sec:scenario}, we introduced independent AWGN to each beam, with an average signal-to-noise-ratio (SNR) of 10 dB.

\begin{figure}[!htbp]
    \centering
    \includegraphics[width=.9\linewidth]{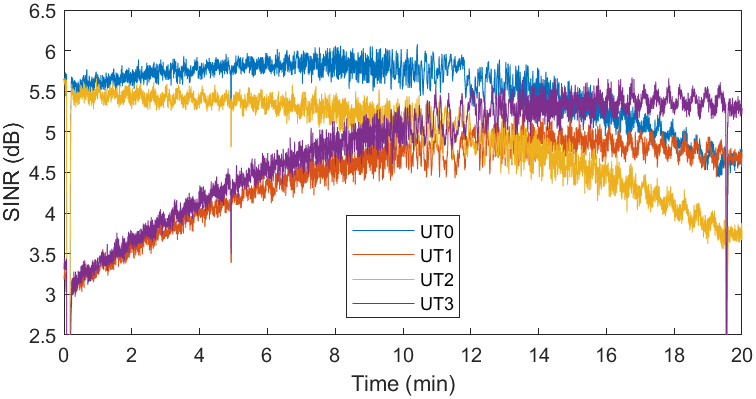}
    \caption{Measured SINR per UT. All impairments are applied. Precoding OFF.}
    \label{fig:results_precoding_off_all_ut}
\end{figure}

We first corroborated that the synchronization chain in the receivers can track the time-varying impairments.
\Cref{fig:results_precoding_off_all_ut} depicts the signal-to-interference-plus-noise-ratio (SINR) measured by each UT when all the impairments are applied in the channel emulator and the precoding is not activated in the GW.
Different effects can be observed, for instance, the time-varying gain of the channel, including the ripple provided by the realistic emulated orbit in the channel emulator.
There are also noticeable perturbations around the center of the plots, corresponding to the changes of sign in the Doppler shifts.
On the other hand, the pronounced variations of SINR at the beginning and end of the plots are artifacts due to the switching times for setting up the scenario in the channel emulator. 

\begin{figure}[!htbp]
    \centering
    \includegraphics[width=.9\linewidth]{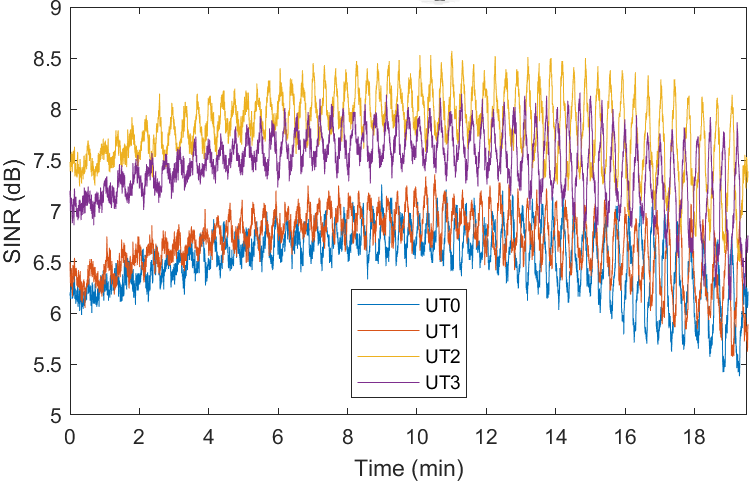}
    \caption{Measured SINR per UT with precoding ON, only variable H and AWGN are applied (baseline).}
    \label{fig:results_precoding_on_H_all_ut}
\end{figure}

Next, we assessed the effect of each of the impairments individually on the precoding performance.
We considered the baseline to be the sub-scenario where only the time-varying channel matrix and AWGN are applied, which provides an upper bound for the precoding gain in terms of SINR increment.
Then we applied each of the other impairments, once at a time, in addition to the time-varying H and noise, and finally the complete scenario (all impairments together). 
\Cref{fig:results_precoding_on_H_all_ut} illustrates the results when precoding is activated for the baseline, with an evident improvement of the perceived SINR compared to the non-precoded case.
Averaging across the UTs, the SINR has an increment that varies between 1.8 and 2.6 dB during the experiment.

\begin{figure}[!htbp]
    \centering
    \includegraphics[width=.9\linewidth]{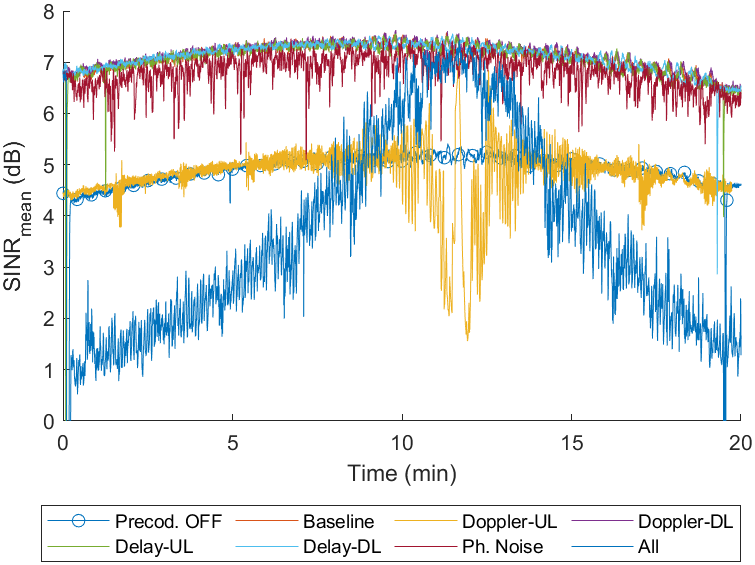}
    \caption{Effect of the different impairments on the precoding performance. Measured SINR averaged across UTs for each tested condition.}
    \label{fig:results_compare_impairments}
\end{figure}

\Cref{fig:results_compare_impairments} shows the measured SINR with precoding, averaged across the four UTs, for each of the tested conditions.
It also includes the SINR without precoding (OFF) for comparison.
The results confirm that the impairments in the downlink (different Doppler shifts and delays) and the delay in the uplink, which is the same for each of the beams, do not affect precoding.
The experiment with independent phase noises in the payload shows that precoding is capable of tracking those phase variations to some extent, although its performance is not optimal because the precoding matrix is only updated at a limited rate.
On the other hand, the different Doppler shifts in the uplink appear as the most critical impairment, without any precoding gain for most of the time, and an oscillatory behaviour of the SINR near the time when the frequency shifts change sign.
Moreover, the joint effect of multiple non-idealities (combining all impairments) is not simply additive and significantly limits the effectiveness of precoding.
In fact, the resulting SINR remains below the achievable value in the non-precoded case, dropping to as low as 1 dB ($>$3~dB loss).
However, the precoding performance improves and achieves almost the baseline results when the satellite passage is around the zero-differential-Doppler interval.
This highlights the complexity of mitigating all MEO impairments simultaneously and underscores the need for robust compensation mechanisms.

\begin{figure}[!htbp]
    \centering
    \includegraphics[width=.9\linewidth]{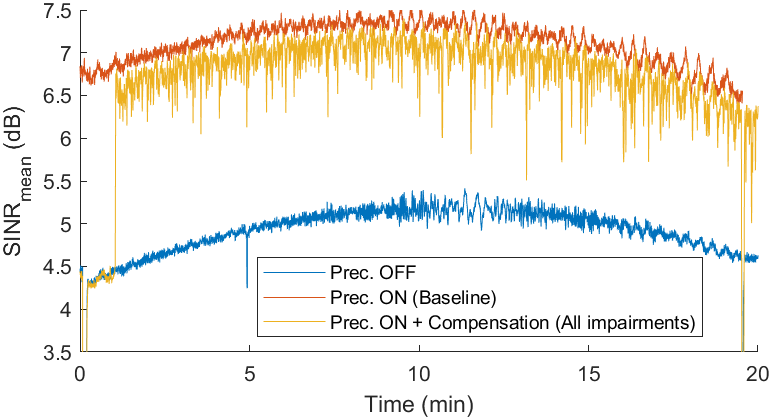}
    \caption{Validation of the effectiveness of the phase compensation technique on the precoding performance. Measured SINR averaged across UTs for each tested condition.}
    \label{fig:results_test_compensation}
\end{figure}

Finally, \Cref{fig:results_test_compensation} demonstrates the effectiveness of the implemented compensation technique, enabling an improvement in precoding performance for most of the satellite passage.
It obtains an SINR very close to the precoding baseline scenario (i.e., what can be obtained with precoding in the absence of other impairments), with a difference of less than 0.5 dB in most of the experimentation time. 

Future work will investigate the optimization of the compensation mechanism to further improve its performance.
We will also look into the emulation of efficient and flexible on-board beamforming techniques focusing on DRAs \cite{Palisetty2023MEO,garces2025egerton,vasquezperalvo2023flexible,vasquezperalvo2024ojcs}.

\section{Conclusions}
\label{sec:conclusions}

This paper has presented the design and validation of an SDR-based in-lab testbed for assessing precoding performance in MEO multibeam satellite systems under realistic conditions. By incorporating accurate orbital models, antenna patterns, and key physical-layer impairments, the setup enables a comprehensive evaluation of the challenges posed by MEO dynamics. Experimental results show that while some impairments have a limited impact on precoding, uplink differential Doppler and independent payload phase noise can significantly degrade its performance. The proposed sample-based phase compensation loop effectively mitigates these effects, recovering most of the precoding gain and approaching the baseline performance with less than 0.5~dB loss.

\section*{Acknowledgment}
The authors would like to thank J. Krause, J. Grotz, and S. Andrenacci from SES, Luxembourg, for their helpful advice on various practical aspects, and N. Mazzali and P.-D. Arapoglou, technical officers from ESA supervising the Livesat-CCN activity.
The authors also thank the rest of Univ. of Luxembourg's team that participated in the Livesat project: J. Querol, N. Maturo, E. Lagunas, R. Palisetty, G. Eappen, H. Chaker, Z. Abdullah, W.A. Martins, and A. Haqiqtnejad. 

\bibliographystyle{IEEEtran}
\bibliography{references.bib}

\end{document}